\documentclass[twocolumn,showpacs,preprintnumbers,amsmath,amssymb]{revtex4}
\usepackage{dcolumn}
\usepackage{bm}
\usepackage[dvipdf]{graphicx}
\DeclareGraphicsExtensions{.jpg,.pdf,.mps,.png,.eps,.ps,.EPS}
                           \begin{document}
\def\be{\begin{equation}}
\def\ee{\end{equation}}
\def\bc{\begin{center}} 
\def\ec{\end{center}}
\def\bea{\begin{eqnarray}}
\def\eea{\end{eqnarray}}

\title{ Loop structure of the Internet at the Autonomous System Level}
\author{Ginestra Bianconi$^1,2$, Guido Caldarelli$^{3,4}$ and Andrea Capocci$^{4  }$}
\affiliation{$^1$The Abdus Salam International Center for Theoretical Physics, Strada Costiera 11, 34014 Trieste, Italy \\
$^2$ INFM, UdR Trieste, via Beirut 2-4, 34014, Trieste,Italy\\
$^3$INFM UdR Roma1 and Dipartimento di Fisica Universit\'a ``La Sapienza'',
P.le A. Moro 2, 00185 Roma, Italy \\
$^4$ Centro Studi e Ricerche E. Fermi, Compendio Viminale, Roma, Italy} 

\begin{abstract}
       
We present here a study of the clustering and loops in the
graph of Internet at the Autonomous Systems level.
 We show that, even if the whole structure is changing with time, the statistical distributions of loops of order 3,4,5 remain stable during the
evolution. 
Moreover we will bring evidence that the Internet graphs  show characteristic Markovian signatures, since its structure is very well described by the two point correlations between the degrees of the vertices. This indeed prove that the Internet belong to a class of network in which the two point correlation is sufficient to describe all their local (and thus global) structure.
Data are also compared to  present Internet models.
\end{abstract}
\maketitle

In the last five years the physics community has started to look at the Internet\cite{Internet_book} as a beautiful example of a complex system with many
degrees of freedom resulting in global scaling properties.
The Internet in fact can be  described as a network, with vertices and edges 
representing respectively Autonomous Systems (AS) and physical lines
connecting them. 
Moreover it has been shown \cite{Faloutsos,calda} that it belongs to
the wide class of scale-free networks \cite{RMP,Doro_book}
emerging as the underline structure of  a variety of real complex systems. 
But, beside the common scale-free connectivity distribution, what distinguish networks as different as the social networks of interactions and the technological networks as for example  the Internet?
                                                                                Researchers have then started to characterize further the networks introducing different topological quantities beside the degree distribution exponent.
Among those, the clustering coefficient $C(k)$\cite{Hierarchical} and the average nearest neighbor degree $k^{nn}(k)$ of a vertex as a function of its degree $k$ \cite{Vespignani1,Vespignani2}.
In particular, measurements in Internet yield $C(k)\sim k^{-\mu}$ with $\mu\simeq{0.75}$
\cite{Vespignani_ps} and $k^{nn}~\sim k^{-\nu}$ with
$\nu\simeq0.5$ \cite{Vespignani_ps}. A two-vertex degree
anti-correlation has also been measured \cite{Maslov}.
Accordingly, Internet is said to display disassortative
mixing \cite{Newman_mixing}, because nodes prefer to be linked to peers
with different degree rather than similar. This situation is opposed to that in social networks where we observe the so-called assortative mixing.

Moreover, the modularity of the Internet due to the
national patterns has been studied by measuring the slow
decaying modes of a diffusion process defined on it \cite{Maslov_dm}.
Recently, more attention has been devoted to network motifs
\cite{UAlon,Milo}, i.e. subgraphs appearing with a frequency larger than  
that observed in maximally random graphs with the same degree sequence.
Among those, the most natural class includes
loops\cite{Loops,Guido_loops,Loop_rev,Avraham}, closed paths of various lengths that visit each node only once.
Loops are interesting because they account for the 
multiplicity of paths between any two nodes. Therefore, they encode  
the redundant information in the network structure. 

In this paper we will present the data of the scaling of the loops of length $h\le 5$ in the Internet graph and we will show that this scaling is very well reproduced by the two points correlation matrix between the degrees of linked pair of vertices.
This allow us to suggest that the Internet is ``Markovian'', i.e.  correlations of order higher than two are negligible.
In the paper we then study the structure of the graph in the two point correlation assumption with the goal of characterizing the cycle structure of the Internet and defining an upper limit of the scaling of the number of loops with the system size valid for all possible lengths of the loops.

\begin{figure}
\includegraphics[width = 60 mm, height = 60 mm ]{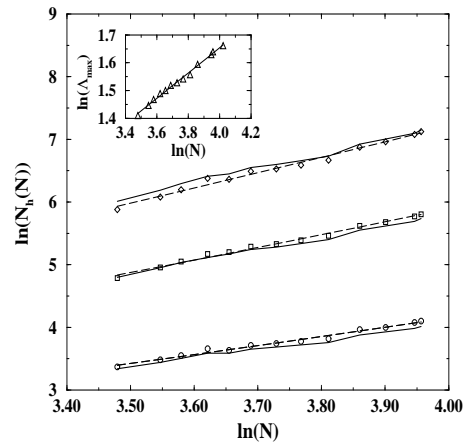}
\caption{\label{AS.fig}Number of $h$-loops $N_h$ as a function of the system size $N$  shown with empty symbols for loops of length 3,4,5 (circles, squares and diamonds). In solid line we report the first order approximation and in dashed line the  power-law fit to the data. In the Inset we report the logarithm of the largest eigenvalue of the matrix $C$ as a function of the system size. }
\end{figure}

To measure the number of loops in an undirected network we  
 consider its symmetrical adjacency matrix $\{a_{ij}\}$, with
$a_{ij} =1$ if $i$ and $j$ are connected and $a_{ij}=0$ otherwise. 
If no loops (self-link in a vertex) are present, i.e.
$a_{ii}=0$ for all $i$, the number of loops of length  $h$ is given by
a dominant term of the type Trace$(a^h)/h$ that counts the total number of paths of length $h$ minus all the contributions coming from intersecting paths.
For $h=3$ this terms are absent and the total number of loops $N_3$ of length $h=3$ is given by
\be
N_3=\frac{1}{6}\sum_i (a^3)_{ii}
\label{n3}
\ee
In the case of shorts loops $h\le 5$ these terms can be easily evaluated and give the expressions  for the total number of loops of size $h=4,5$, $N_4,N_5$ \cite{Loops}
\bea
N_4=\frac{1}{8}\left[\sum_i (a^4)_{ii}-2\sum_i(a^2)_{ii} (a^2)_{ii}+\sum_i (a^2)_{ii}\right] \nonumber \\
N_5=\frac{1}{10}\left[\sum_i (a^5)_{ii}-5\sum_i(a^2)_{ii} (a^3)_{ii}+5\sum_i (a^3)_{ii}\right].
\label{n}
\eea
To measure the actual scaling in Internet at the AS level, 
we  used Eqs.$(\ref{n3})-(\ref{n})$.
The data of the Internet at the Autonomous System level are collected by the 
University of Oregon Route Views Project and made available by the 
NLANR (National Laboratory of Applied Network Research). 
The subset we used in this manuscript 
are mirrored at  COSIN web page http://www.cosin.org.
We considered 13 snapshots of the Internet network at the AS level at
different times starting from November 1997 (when $N=3015$) toward
January 2001 ($N=9048$). 
Throughout this period, the degree distribution is a power-law
with a nearly constant exponent $\gamma \simeq 2.22(1)$.
Using relations $(\ref{n3})$, $(\ref{n})$, we measure
$N_h(t)$ for $h=3,4,5$ in the Internet at different times,
corresponding to different network size.
We observe in figure \ref{AS.fig} that the data follow a scaling of the type 
\be
\label{powlaw}
N_h(N)\sim N^{\xi(h)}
\end{equation}
with the $\xi(h)$ exponents reported in table $\ref{xi_table}$.
 
To model the Internet means to find a class of networks defined by a stochastic algorithm that share the main characteristic of the Internet graph. Consequently we suppose that the real Internet graphs belong to a certain ensemble of graphs and it is actually a realization of it.
Supposing one knows this ensemble in order to evaluate the number of loops one
 theoretically  would need  to know the entire probability distribution for each element of the adjacency matrix, i.e. the probability distribution $P(a_{11},\dots a_{1,N},\dots a_{N,1}\dots a_{N,N})$.
Lets make the assumption that  the probability for a set of $h$-nodes to be connected depends only on the connectivities.
The zero order  approximation to Eqs.$(\ref{n3})-(\ref{n})$ would be then to assume that the connectivity of the nodes are completely uncorrelated and then
the formula for calculation of the loops of size $h$, would be \cite{poli2}
\bea
N_h^{(1)}=\frac{1}{2h}\left[\frac{\sum_k k (k-1)P(k)}{<k>}\right]^h,\nonumber \\
\eea
Given a distribution $P(k)~k^{-\gamma}$ with a cutoff at $k_c=N^{1/\chi}$
we get the scaling prediction Eq. $(\ref{powlaw})$
with
$\xi(h)=h (3-\gamma)/{\chi},$
in the relevant case $2<\gamma<3$.
In the special case of a uncorrelated graph with $\gamma=3$
we obtain the scaling behavior 
$
N_h(N)\sim\left( \log(N)\right)^{\psi(h)},$

with $\psi(h)=h$.
Interesting enough the same calculation is exactly valid also in a  Barab\'asi-Albert\cite{BA} network which is a off-equilibrium network but with zero correlations \cite{Loops}.
We need to observe that the fact itself that in the Internet data the exponent $\chi$ follows 
\be
\frac{1}{\chi}=\frac{1}{\gamma-1}
\label{psi.eq}
\ee
indicates that the network is strongly correlated, in fact for uncorrelated networks we would expect $1/\psi=1/2$ \cite{poli1,FS_vesp}.

The real exponents $\xi(h)$ as expected depend on $h$,  but unfortunately they significantly differ from the
zero order approximation values
 $\xi(h)=h(3-\gamma)/\chi$ with $\chi$ given by Eq.$(\ref{psi.eq})$ for and  $\gamma\simeq 2.22$ (see  table \ref{xi_table}).
So, the correlation nature of the Internet cannot be neglected when one looks at the scaling of the loops in the network.

\begin{figure}
\includegraphics[width = 60 mm, height = 60 mm ]{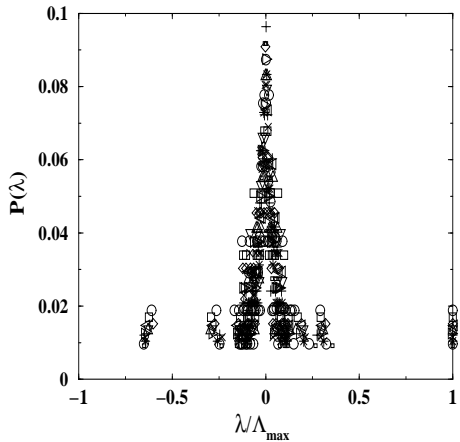}
\caption{\label{spectra.fig}
The rescaled spectra of the matrix $C$ calculated over the $13$ snapshots of the Internet under study.
}
\end{figure}

The first order approximation for Eqs. $(\ref{n3})-(\ref{n})$  consists on
taking into account that the connectivity of the nodes are correlated.
In order to calculate the number of small loops in the network one can  approximate $N_h\sim\mbox{Trace}(a^h/2 h)$.
In a loop all the nodes are equivalent, having two links,  fixed a direction on the loop one link is used to reach the given node an the other link to reach the subsequent.
The probability that a  node of degree $k_1$,already part of the loop, is connected to a successive node of degree $k_2$ is given by $(k_1-1)P(k_2|k_1)$ since we can decide to follow one of its remaining $k_1-1$ nodes. (In our notations  $P(k|k')$ indicates  the probability that following one link starting at node $k'$ one reaches a node with connectivity $k$).
 Consequently, the number of loops of size $h$ in this first order approximation are given by
\bea
N_h^{(2)}=\frac{1}{2h}\mbox{Trace}(C^h)
\label{SOA.eq}
\eea
where the matrix  $C$ is defined as
\be
C_{k,k'}= (k'-1)P(k|k').
\ee
Of course for higher order loops it will be not possible to neglect the contributions of intersecting paths, but still Eq. $(\ref{SOA.eq})$ would provide an upper limit to the behavior of $N_h(N)$.
In Fig. $\ref{AS.fig}$ we compare the real data with the first order approximation given by Eqs.$(\ref{SOA.eq})$.
It is clear that this approximation capture most of the cycle structure, at least for small value of $h$.
Since we observe this peculiar characteristic of the Internet graphs is worth to look at the structure of the matrix $C$.
Indeed the matrix $C$ is characterized by a spectra in which there with  eigenvalues $\lambda$ which scale as
\be
\lambda(N)\sim N^{\theta}
\ee
where $\theta=0.47\pm0.01$.
In Fig. $\ref{spectra.fig}$ we show  how this spectra scales for the different snapshots of the Internet at the Autonomous System Level.
The largest eigenvalue $\Lambda_{max}(N)$ is the one of  much interest to us in this letter since it is responsible for the behavior of $N_h$ at large $N$.Indeed we can estimate an upper limit for the scaling of the loops of generic length $h$ with the system size, i.e.
$
N_h^{(2)}\leq O\left({\Lambda_{max}^h}/{2h}\right)$
where is scaling is supposed to be valid until  $h\ll h^*$ where some
arguments support the scaling  $h^*\sim N^{\frac{(3-\gamma)}{2}}$ for random scale-free graphs\cite{Large} and $h^*\sim N^{1/(\gamma-1)}$ for correlated graphs \cite{Avraham} (see for the behavior of the number
loops at large $h$ in regular random graphs \cite{Marinari}).
In order to fully characterize the cycle structure of the Internet is then 
natural to study the structure of the eigenvector associated to the largest eigenvalue.
For this vector $u_k$ also we observe a scaling behavior
\be
u_k=k^{\alpha}f(k/k_c) 
\ee
where 
$f(x)= 1 $ for $ x\ll1$ and 
$f(x)=x^{\beta}$ for $x\gg1$,
with $\alpha=-2.50\pm0.05$ and $\beta=3.10\pm0.05$.

To make a comparison between the real data and the model present in the literature at the moment we consider the Fitness model\cite{Fitness} and the Generalized Network Growth Model (GNG)\cite{GNG} and the Competition and Adaptation Model\cite{Boguna} with (D) and without(ND) distance constraints.
The fitness model  has indeed $\gamma=2.255$ and the GNG model has a power-law exponent that depends on the intrinsic parameter $p$ $\gamma(p)=2+p/(2-p)$.
In order to compare networks with a similar mean degree ($<k>\in (3.4-4.0)$\cite{nota} for
the Internet), we consider the fitness model with $m=2$ ($<k>=2m=4$)
and the GNG model with parameter $p=0.5$ ($<k>=2/p=4$) and $p=0.6$ 
($<k>=2/p=3.33$).All models present not trivial correlations of the nodes as can be seen by observing the $C(k)$ and $k^{nn}(k)$ functions.

In table $\ref{xi_table}$ we compare the $\xi(h)$ exponents of the  real data
with the exponents numerically calculated for the considered models. While $\xi(h)$ grows almost linearly with $h$ as expected we observe that the D and ND models seems to best reproduce the data. 
\begin{table}[h] 
\begin{center}
\begin{tabular}{|c|c|c|c|}
\hline
System & $\xi(3)$ & $\xi(4)$ & $\xi(5)$ \\
\hline
AS & $1.45 \pm 0.07$ & $2.07 \pm 0.01$ & $2.45 \pm 0.08$ \\
ZOA & $2.26\pm0.06$ & $3.15\pm 0.07$ & $3.94\pm0.09$ \\
FOA & $1.34\pm0.03$ & $1.86\pm 0.04$ & $2.25\pm 0.05$\\
Fitness & $0.59 \pm 0.02$ & $0.86 \pm 0.02$ & $1.10 \pm 0.02$ \\
GNG (p=0.5) & $0.53 \pm 0.03$ & $0.72 \pm 0.03$ & $0.96 \pm 0.02$ \\
GNG (p=0.6) & $0.53 \pm 0.03$ & $0.74 \pm 0.03$ & $0.99 \pm 0.02$ \\
D& $1.60\pm 0.01$ & $2.20\pm 0.03$ & $2.70\pm0.03$\\
ND & $1.59\pm 0.03$ &$2.11\pm 0.03$ & $2.64 \pm0.03$\\
\hline
\end{tabular}
\end{center}
\caption{The exponent $\xi(n)$ for $n=3,4,5$ as defined in
equation (\ref{powlaw}) for real data,in the zero order approximation (Z0A) and in the first order approximation (FOA),and for network models.}
\label{xi_table}
\end{table}
\begin{figure}
\includegraphics[width = 60 mm, height = 60 mm ]{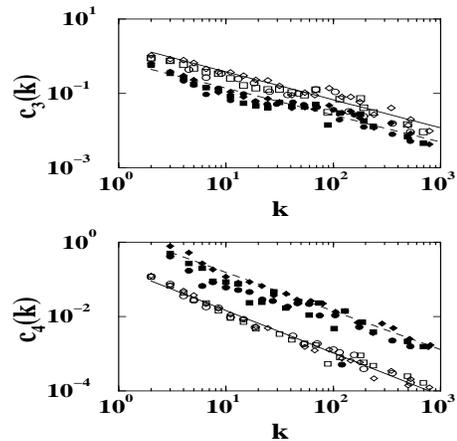}
\caption{\label{C3C4.fig}
The clustering coefficients $c_3(k)$ and $c_4(k)$ in Internet for the data 
of    November '97 (circles), January
'99 (squares) and  January '01 (triangles). In filled symbols the same results obtained in the first approximation assumption.In solid and dashed lines we indicate the power-law fit to the data and to the first order approximation results respectively.
}
\end{figure}

Following \cite{Guido_loops}, we also measured the clustering coefficients $c_{3,i}$ and
$c_{4,i}$ as a function of the connectivity $k_i$ of node $i$ for all $i$'s.
In particular, $c_{3,i}$ is the usual clustering coefficient $C$, i.e. the number of 
triangles including node $i$ divided by the number of possible triangles $k_i(k_i-1)/2$.

Similarly, $c_{4,i}$ measures the number of quadrilaterals passing through node $i$ 
divided by the number of possible quadrilaterals $Z_i$. 
This last quantity is the sum of all possible primary quadrilaterals $Z_i^p$ 
(where all vertexes are nearest neighbors of node $i$) 
and all possible secondary quadrilaterals $Z_i^s$ 
(where one of the vertexes is a second neighbor of node $i$).
If node $i$ has $k_i^{nn}$ second neighbors, 
$Z_i^p=k_i(k_i-1)(k_i-2)/2$ and $Z_i^s=k_i^{nn}k_i(k_i-1)/2$.
In Fig. \ref{C3C4.fig} (a) we plot $c_3(k)$, $c_4(k)$ 
for the Internet data at three different times (November 1997, 
January 1999 and January 2001) showing that the behavior of $c_3(k)$ 
and $c_4(k)$ is invariant with time and scales as 
\begin{equation}
c_h(k)\sim k^{-\delta(h)}
\label{delta.eq}
\end{equation}
with $\delta(3)=0.7(1)$ and $\delta(4)=1.1(1)$.

In Fig. $\ref{C3C4.fig}$, we compare the behavior of $c_3(k)$ and $c_4(k)$
in real Internet data  with  the first order approximation (FOA) results.
Again we observe that the first-order approximation results are quite satisfactory reinforcing our thesis that to explain the loop  structure of the Internet it is sufficient to stop at this order. However, the behavior of $c_3(k)$ and $c_4(k)$ cannot be explained just looking at the largest eigenvalues of the $C$ matrix but one has to consider the entire spectra.
For completeness we also considered the behavior of  the clustering coefficients $c_3(k)$ and $c_4(k)$ in Internet models Table \ref{esponenti_delta}. We observe that while in the (D) and (ND) models there are large deviations form the  scaling $\ref{delta.eq}$ these models seems in general to capture better the cycle structure of the Internet respect to the other non ad-hoc models we have considered here.
\begin{table}[b] 
\begin{center}
\begin{tabular}[c]{|c|c|c|}
\hline
System & $\delta(3)$ & $\delta(4)$ \\
\hline
AS & $0.75\pm 0.05$& $1.13\pm 0.05$ \\
SOA &$ 0.70\pm0.05$ & $1.00\pm 0.05$ \\
Fitness & $0.67 \pm 0.01$ & $0.99 \pm 0.01$ \\
GNG (p=0.5) & $0.32 \pm 0.02$ & $1.68 \pm 0.03$ \\
GNG (p=0.6) & $0.27 \pm 0.02$ & $0.93 \pm 0.01$  \\
D &$0.3\pm 0.2$ &$0.8 \pm 0.2$\\
ND & $0.6 \pm 0.2$ &$1.0 \pm 0.2$\\
\hline
\end{tabular}
\end{center}
\caption{The exponent of the clustering coefficient $c_3(k)$ and
$c_4(k)$ as measured from Internet data as a result of the SOA and from simulations of Internet models.}
\label{esponenti_delta}
\end{table}

In conclusion, we computed the number $N_h(t)$ of $h$-loops of 
size $h=3,4,5$ in the Internet at the Autonomous System level and  the generalized clustering coefficients around individual nodes
as a function of nodes degrees.
We have observed that this evolving  network has a structure of the loops that
is well captured by the two point correlation matrix. Indeed it seems that the Internet is ``Markovian'' in the sense that is not necessary to study a correlations function of more that two points, at least to explain the cycle structure. 
For this reason we have characterized the   correlations matrix $C_{k,k'}=(k'-1)P(k|k')$ studying its spectra  and the structure of the  eigenvector associated with the maximal eigenvalue.
Finally we  have  compared these results with the behavior of the same quantities $N_h(N)$ and $c_h(k)$ in the fitness model, in the GNG model  and the D, ND models, a
a chosen subset of the available Internet models present in the literature, finding that the ad-hoc D, ND model seems to capture better the cycle structure of the Internet.

The authors are grateful to Uri Alon, Shalev Itzkovitz Matteo Marsili and Yi-Cheng Zhang  
for useful comments and discussions and to the authors of Ref. \cite{Boguna} for suggesting to measure the cycle structure of their model.
This paper has been financially supported by the Swiss National Foundation, under grant 
no. 2051-067733.02/1, and by the European Commission - FET Open project COSIN IST-2001-33555 and IP project DELIS and EVERGROW.


\begin{thebibliography}{}
\bibitem{nota}
Indeed the Internet show an accelerated growth \cite{Faloutsos,Acce_Growth} that is not taken into account into this model. Indeed the acceleration is rather small $<k>\sim N^{\alpha}$ with $\alpha=0.01$ and has been neglected in these models.
\end{thebibliography}
\end{document}